\documentclass{article}
\usepackage[T1]{fontenc} % add special characters (e.g., umlaute)
\usepackage[utf8]{inputenc} % set utf-8 as default input encoding
\usepackage{ismir,amsmath,cite,url}
\usepackage{graphicx}
\usepackage{color}

\usepackage{cleveref}
\usepackage{booktabs}

\title{
LakhNES:
Improving multi-instrumental music generation \\
with cross-domain pre-training
}

\multauthor
{Chris Donahue$^1$ \hspace{10mm}
Huanru Henry Mao$^2$ \hspace{10mm}
Yiting Ethan Li$^2$ \hspace{10mm}
}
{
\bfseries{
Garrison W. Cottrell$^2$ \hspace{1cm} Julian McAuley$^2$
}
\\
$^1$ Department of Music, UC San Diego \hspace{10mm}
$^2$ Department of Computer Science, UC San Diego
}
\def\authorname{Chris Donahue, Huanru Henry Mao, Yiting Ethan Li, Garrison W. Cottrell, Julian McAuley}

\begin{document}

\maketitle
\begin{abstract}
%As such, their success remains elusive in domains where (relatively) less data is available, such as the multi-instrument setting. 
%By capturing scores for the four-voice ensemble of an early video game sound synthesis chip, we construct a dataset of suitable size for effectively training such a model. 
We are interested in the task of generating multi-instrumental music scores.
The \emph{Transformer} architecture has recently shown great promise for the task of piano score generation---here we adapt it to the multi-instrumental setting.
Transformers are complex, high-dimensional language models which are capable of capturing long-term structure in sequence data, but require large amounts of data to fit. 
Their success on piano score generation is partially explained by the large volumes of symbolic data readily available for that domain. 
We leverage the recently-introduced \emph{NES-MDB} dataset of four-instrument scores from an early video game sound synthesis chip (the NES), which we find to be well-suited to training with the Transformer architecture.
To further improve the performance of our model, we propose a pre-training technique to leverage the information in a large collection of heterogeneous music, namely the Lakh MIDI dataset. 
Despite differences between the two corpora, we find that this transfer learning procedure improves both quantitative and qualitative performance for our primary task.
\end{abstract}

\section{Introduction}\label{sec:introduction}

In this paper, we extend recent results for symbolic piano music generation~\cite{huang2019music} to the 
multi-instrumental setting. 
Both piano and multi-instrumental music are \emph{polyphonic}, where multiple notes may be sounding at any given point in time.
However, the generation of multi-instrumental music presents an additional challenge not present in the piano domain: 
%JULIAN: This argument is only okay. There could be different methods of generation that don't depend on this mapping, e.g. generating one instrument at a time conditioned on the others. This seems a challenge with a specific solution rather than a general challenge.
%producing a mapping which assigns each note in the score to a suitable instrument. 
handling the intricate interdependencies between multiple instruments. 
% and unique sound-producing capabilities of multiple instrument voices
Another obstacle for the multi-instrumental setting is that there is less data available than for piano, 
making it more difficult to train the types of powerful generative models used in~\cite{huang2019music}. 
%Here we offer solutions to both problems.

Until recently, 
music generation methods struggled to capture two rudimentary elements of musical form: long-term structure and repetition. 
Huang et al.~\cite{huang2019music} demonstrated that powerful neural network \emph{language models}, i.e.,~models which assign likelihoods to sequences of discrete tokens,
could be used to generate classical piano music containing these elusive elements. 
%To our ears, the piano music generated from their \emph{Music Transformer} model represents the most compelling computer-generated music to date. 
In order to adapt this method to the multi-instrumental setting
we incorporate instrument specification directly into our language-like music representation.
However, 
this strategy alone may be insufficient to generate high-quality multi-instrumental music, 
as the results of~\cite{huang2019music} also depend on access to large quantities of piano music.

To begin to address the data availability problem, 
we focus on an unusually large dataset of multi-instrumental music. 
The \emph{Nintendo Entertainment System Music Database} (NES-MDB)~\cite{donahue2018nesmdb} contains $46$ hours of \emph{chiptunes}, 
music written for the four-instrument ensemble of the NES (video game system) sound chip. 
This dataset is appealing for music generation research not only for its size but also for its structural homogeneity---all of the music is written for a fixed ensemble. 
It is, however, smaller than the $172$ hours of piano music in the \emph{MAESTRO Dataset}~\cite{hawthorne2018enabling} used to train Music Transformer.

The largest available source of symbolic music data is the \emph{Lakh MIDI Dataset}~\cite{raffel2016lakh} which contains over $9000$ hours of music. 
This dataset is structurally heterogeneous (different instruments per piece) making it challenging to model directly.
However, 
intuition suggests that we might be able to benefit from the musical knowledge ingrained in this dataset to improve our performance on chiptune generation.
Accordingly, 
we propose a procedure to heuristically map the arbitrary ensembles of music in Lakh MIDI into the four-voice ensemble of the NES. 
We then pre-train our generative model on this dataset, 
and fine-tune it on NES-MDB. 
We find that this strategy improves the quantitative performance of our generative model by $10$\%. 
Such transfer learning approaches are common practice in state-of-the-art natural language processing~\cite{devlin2018bert,radford2018language}, 
and here we develop new methodology to employ these techniques in the music \emph{generation} setting (as opposed to analysis~\cite{choi2017transfer}).
%and we believe the broader music generation community can benefit from this strategy to overcome data availability issues.

We refer to the generative model pre-trained on Lakh MIDI and fine-tuned on NES-MDB as \emph{LakhNES}. 
In addition to strong quantitative performance, 
we also conduct multiple user studies indicating that LakhNES produces strong qualitative results. 
LakhNES is capable of generating chiptunes from scratch, 
continuing human-composed material, 
and producing melodic material corresponding to human-specified rhythms.\footnote{
Sound examples: \url{https://chrisdonahue.com/LakhNES}
Code/data: \url{https://github.com/chrisdonahue/LakhNES}} 

\section{Related work}
\label{sec:related}

Music generation has been an active area of research for decades. 
Most early work involved manually encoding musical rules into generative systems or rearranging fragments of human-composed music;
see~\cite{nierhaus2009algorithmic} for an extensive overview. 
Recent research has favored machine learning systems which automatically extract patterns from corpora of human-composed music. 

Many early machine learning-based systems focused on modeling simple \emph{monophonic} melodies, i.e.,~music where only one note can be sounding at any given point in time~\cite{todd1989connectionist,mozer1994neural,eck2002finding}.
More recently, research has focused on \emph{polyphonic} generation tasks.
Here, most work represents polyphonic music as a \emph{piano roll}---a sparse binary matrix of time and pitch---and seeks to generate sequences of individual piano roll timesteps~\cite{boulanger2012modeling,johnson2017generating} or chunks of timesteps~\cite{yang2017midinet}. 
Other work favors an \emph{event-based} representation of music, where the music is flattened into a list of musically-salient events~\cite{simon2017performance,mao2018deepj,huang2019music}. 
None of these methods allow for the generation of multi-instrumental music.
%, as they provide no mechanism for mapping the generated polyphonic scores to individual instruments. 

Other research focuses on the multi-instrumental setting and seeks to provide systems which can \emph{harmonize} with human-composed material~\cite{allan2005harmonising,huang2016counterpoint,hadjeres2017deepbach,yan2018part}.
Unlike the system we develop here, these approaches all require complex inference procedures to generate music without human input. 
Recent work~\cite{dong2018convolutional,roberts2018hierarchical,dong2018musegan} attempts multi-instrumental music generation from scratch, but these methods are limited to generating fixed lengths, unlike our method which can generate arbitrarily-long sequences. 
There is also music generation research that operates on the audio domain~\cite{donahue2018adversarial,dieleman2018challenge}, 
though this work is largely unrelated to symbolic domain methods.
The work described in this paper is methodologically similar to MuseNet~\cite{payne2019musenet}, which was concurrent with our work.

\section{Datasets and task}

The \emph{NES Music Database} (NES-MDB)~\cite{donahue2018nesmdb} consists of approximately $46$ hours of music composed for the sound chip on the Nintendo Entertainment System. 
This dataset is enticing for research in multi-instrumental music generation because (1) it is an unusually large corpus of music that was composed for a fixed ensemble, and (2) it is available in symbolic format.

\subsection{NES ensemble preliminaries}
\label{sec:nesens}

The ensemble on the NES sound chip has four monophonic instrument voices: two pulse waveform generators (P1/P2), one triangle waveform generator (TR), and one noise generator (NO).\footnote{There is an additional fifth voice capable of waveform playback that the authors of NES-MDB excluded.}
The first three of these instruments are melodic voices: 
typically, TR plays the bass line and P1/P2 are interchangeably the melody and harmony. 
The noise instrument is used to provide percussion.

The various instruments have a mixture of sound-producing capabilities. 
For example, the range of MIDI pitches which P1/P2 can generate is $33$--$108$, while the range of TR extends an octave lower ($21$--$108$). 
The noise channel can produce $16$ different ``types'' of noise which correspond to different center frequencies and bandwidths. 
Each instrument also has a variety of dynamics and timbral attributes. 
It is shown in~\cite{donahue2018nesmdb} that these expressive attributes can be estimated from the score post-hoc, 
and hence we ignore them in this study to focus on the problem of modeling composition rather than expressive performance.

Each chiptune in NES-MDB is stored as a MIDI file, and the constituent MIDI events are quantized at audio rate ($44100$ \emph{ticks} per second). 
Paired with code which synthesizes these MIDI files as NES audio,
the files contain all of the information needed to synthesize the original $8$-bit waveforms.
%It is a unique aspect of this dataset that it is paired with an ensemble with fixed acoustic properties
%It is a unique aspect of this dataset that it is paired with code to synthesize the symbolic music by ensemble for which it was originally composed.

% Paired with code for audio synthesis, NES-MDB includes all of the information needed to synthesize the original NES-generated waveform with sample-accurate precision. 
% %CHRIS: probably remove this paragraph
% To create NES-MDB, the authors first gathered an initial dataset consisting of \emph{Video Game Music} (VGM) files. 
% These files were created by enthusiasts by emulating particular NES games and recording the timing and values of writes made by the system to any memory address associated with the APU.
% However, 
% the VGM format is marred by unnecessary and redundant commands, 
% and it is difficult to interpret musical meaning. 
% To rectify this, the authors of NES-MDB emulated the functionality of the NES APU, converting the unwiedly VGM data into simple MIDI files.

\subsection{Event-based task}
\label{sec:eventbased}

\begin{figure}[t]
    \centering
    \includegraphics[width=0.99\columnwidth]{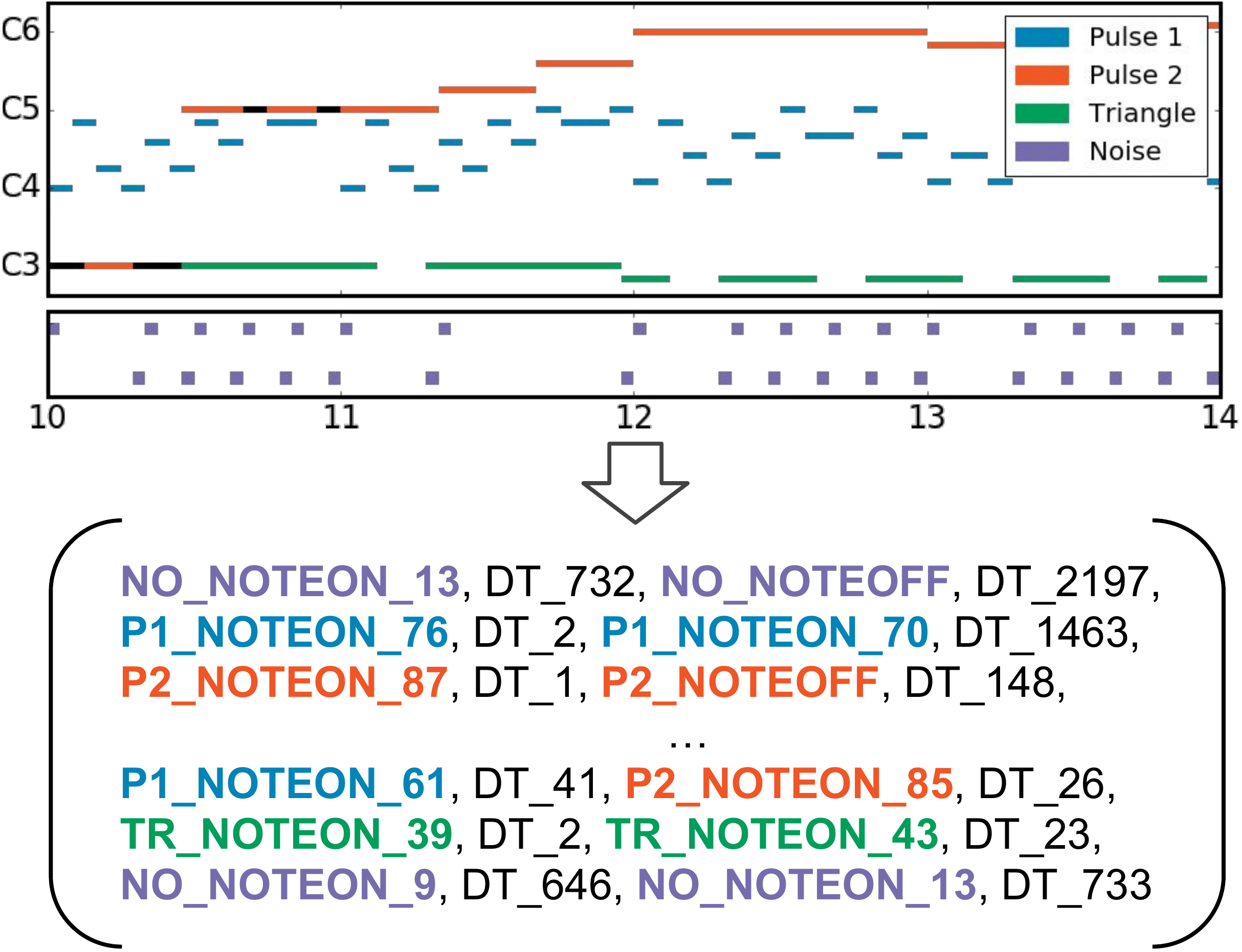}
    \caption{A visual comparison between the piano roll representation of the original NES-MDB paper~\cite{donahue2018nesmdb} (\textbf{top}) and the event representation of this work (\textbf{bottom}). In the piano roll representation, the majority of information is the same across timesteps. In our event representation, each timestep encodes a musically-meaningful change.}
    \label{fig:evtrep}
\end{figure}

In our original work on NES-MDB~\cite{donahue2018nesmdb},
we operated on a \emph{piano roll} representation of the data, i.e.,~the MIDI information decomposed into a sparse grid across time, pitch, and instrument (top of \Cref{fig:evtrep}). 
Because no tempo or beat information exists in the dataset, the authors chose to discretize the time axis at a rate of $24$ timesteps per second. 
This high rate is necessary for capturing nuanced timing information in the scores but results in much of the information being redundant across adjacent timesteps. 
This represents a challenge as long-term dependencies are a barrier to success for sequence modeling with machine learning.

To circumvent these issues, we design an \emph{event-based} representation (bottom of \Cref{fig:evtrep}) similar to that used for single-instument music in~\cite{simon2017performance}. 
Specifically, 
we convert each NES-MDB MIDI file into a time-ordered sequence of events, 
so that every entry in the sequence corresponds to a musically-salient occurrence. 

To handle the rhythmic information, we add time shift ($\Delta{}T$) events which represent time advancing by some discrete number of ticks (each tick is $\frac{1}{44100}$th of a second).
%(there are $44100$ ticks per second as NES-MDB MIDI are quantized at audio sampling rate). 
To keep the number of events in our representation tractable, 
we quantize $\Delta{}T$ events in the real data to fixed gratings. 
We embed the multi-instrumental aspect of our problem directly into this representation by using separate note on/off events for each instrument.
Contemporaneous events are always listed in the following instrument order: P1, P2, TR, NO.
Our final representation consists of $631$ events,
of which about half encode time-related events and half note-related (\Cref{tab:perfrep}). 
Apart from minor timing quantization, this format is a lossless transformation of the original MIDI score.
%We will release the dataset in this format upon publication.

\begin{table}[t]
\centering
\begin{tabular}{lr}
\toprule
Event description & Event ID(s) \\
\midrule
Start or end of sequence & $0$ \\
$\Delta{}T$ for $1$--$100$ ticks (short) & $1$--$100$ \\
$\Delta{}T$ for $100$--$1000$ ticks (medium) & $101$--$190$ \\
$\Delta{}T$ for $>10000$ ticks (long) & $191$--$370$ \\
P1 Note Off/On & $371$--$447$ \\
P2 Note Off/On & $448$--$524$ \\
TR Note Off/On & $525$--$613$ \\
NO Note Off/On & $614$--$630$ \\
\bottomrule
\end{tabular}
\caption{Schematic for our event-based representation of NES-MDB, reminiscent of the one used in \emph{Performance RNN}~\cite{simon2017performance}. The $631$ events in our representation are distributed among time-shift ($\Delta{}T$) events (which allow for nuanced timing), and note off/on events for individual instruments (as in typical MIDI).}
\label{tab:perfrep}
\end{table}

\section{Methodology}

To model the event sequences outlined in the last section, we adopt a \emph{language modeling} factorization.
%, so-called because it originates from natural language processing. 
%JULIAN: Above sentence is a bit verbose
We factorize the joint probability of a musical sequence consisting of $N$ events ($E_1, \ldots{}, E_N$) into a product of conditionals:
\begin{equation}
P(E_1) \cdot P(E_2 \mid E_1) \cdot \ldots \cdot P(E_N \mid E_{1}, \ldots, E_{N - 1}).
\end{equation}
This factorization is convenient because it allows for 
%JULIAN: Could drop the following line
a simple left-to-right algorithm for generating music: 
sampling from the distribution estimated by the model at each timestep (conditioned on previous outputs).
The goal of our optimization procedure is to find a model configuration which maximizes the likelihood of the real event sequences. 
Motivated by the strong results for piano music generation from the recent \emph{Music Transformer}~\cite{huang2019music} approach, 
we also adopt a Transformer~\cite{vaswani2017attention} architecture.

\subsection{Transformer architecture}

The Transformer~\cite{vaswani2017attention} is an attention-based neural network architecture. 
In our context, this means that the model has a mechanism which explicitly biases its predictions based on a subset of musical events that have happened in the past. 
%JULIAN: Sentence below is somewhat garbly
%The mapping which determines the subset of past musical events that the model pays attention when predicting the current event is learned. 
The model's design gives it the ability to learn which subset of past musical events to pay attention to when predicting the current event.
This mechanism may be especially useful for learning patterns of repetition in music across large gaps of time.
%The hope is that this mechanism can capture long-term trends in music, and may be especially useful for learning patterns of repetition in music. 

The original Transformer architecture~\cite{vaswani2017attention} was an encoder-decoder model designed for language translation. 
In this paper, we are only concerned with the decoder portion of the Transformer. 
%JULIAN: Remainder of this paragraph doesn't add much, could be dropped
%The Transformer decoder consists of several layers of processing each with its own attention mechanism. 
%Compared to recurrent neural networks, attention is easier to parallelize during training and excels empirically at modeling long-range dependencies. 
%self-attention has been shown to be easy to parallelize in training and excellent at modeling long-range dependencies due to its smaller maximal path length between distant tokens. 
%For a detailed description of Transformer, we refer the reader to~\cite{vaswani2017attention}.
Our work uses a recent extension of Transformer called Transformer-XL~\cite{dai2019transformer}, 
which is designed specifically to handle longer sequences. 
Transformer-XL builds upon the Transformer architecture by augmenting it with a recurrence mechanism.
%and improved relative positional encoding. 
The recurrence mechanism enables Transformer-XL to use information beyond its training segment by learning how to incorporate recurrent state from previous segments. 
In contrast, the original Transformer is only able to alter its predictions based on the current training segment, hence the available system memory during training is a bottleneck to its ability to learn long-term dependencies.
%This addresses the \textit{context fragmentation} problem, whereby the model is unable to condition on tokens beyond its training window. 
In order to effectively use its recurrent state, Transformer-XL adopts a sophisticated position-aware mechanism so the model can generalize to different amounts of recurrent memory during generation. 

The Music Transformer \cite{huang2019music} is a different Transformer variant that also attempts to tackle long-range dependencies by using a mechanism which reduces the quadratic memory cost of attention, enabling training on longer sequences. 
Although similar in goal to Transformer-XL, its method is orthogonal and could, in theory, be combined with the recurrent mechanism of Transformer-XL. 
For simplicity, we focus on the Transformer-XL architecture as its recurrence mechanism alone is sufficient to learn long-term dependencies. 
Additionally, code to reproduce the Music Transformer method is unavailable.
%Furthermore, the relative positional encoding in Transformer-XL is more sophisticated than the Music Transformer due to the inductive bias built into its use of the sinusoidal positional encoding. 

\subsection{Pre-training}

Transformers are extremely high-dimensional models, and accordingly they can learn effective strategies for extremely large datasets~\cite{radford2018language}. 
One barrier to their application in the music domain is that most symbolic music datasets are either too small or too structurally heterogeneous. 
For example, the popular Bach chorales dataset~\cite{hild1992harmonet} is structurally homogeneous (all chorales have four voices), but small (only $306$ chorales).
In contrast, 
the Lakh MIDI dataset~\cite{raffel2016lakh} is enormous ($175$k songs) but heterogeneous (varying numbers of instruments per piece). 
The NES-MDB dataset we use in this work represents a middle ground (large and structurally-homogenous), but is still substantially smaller than the MAESTRO dataset~\cite{hawthorne2018enabling} used to train Music Transformer ($46$ hours vs. $172$ hours).

We hypothesize that we can improve the performance of our model on our NES music generation task by leveraging the musical information in the larger Lakh MIDI dataset. 
To test this, we propose a two-step procedure.
First, we map each structurally-heterogeneous Lakh MIDI file into one which can be performed by our NES ensemble. 
Then, we pre-train a Transformer on this dataset, and fine-tune this pre-trained model on the NES-MDB dataset. 
Such transfer learning procedures are common methodology in other areas of machine learning~\cite{pan2010survey}, but remain hitherto unexplored in music generation research. 
One possible reason for the lack of investigation into this strategy is that 
mapping music from one domain to another requires careful consideration of musical invariants, 
and hence is less straightforward than analogous methodology for other tasks (e.g.,~language).
We consider this transfer learning protocol
%(and the general idea of employing transfer learning for music generation) 
to be a primary methodological contribution of this work.

\subsubsection{Mapping Lakh MIDI to the NES ensemble}
\label{sec:lakh2nes}

\begin{figure}[t]
    \centering
    \includegraphics[width=0.99\columnwidth]{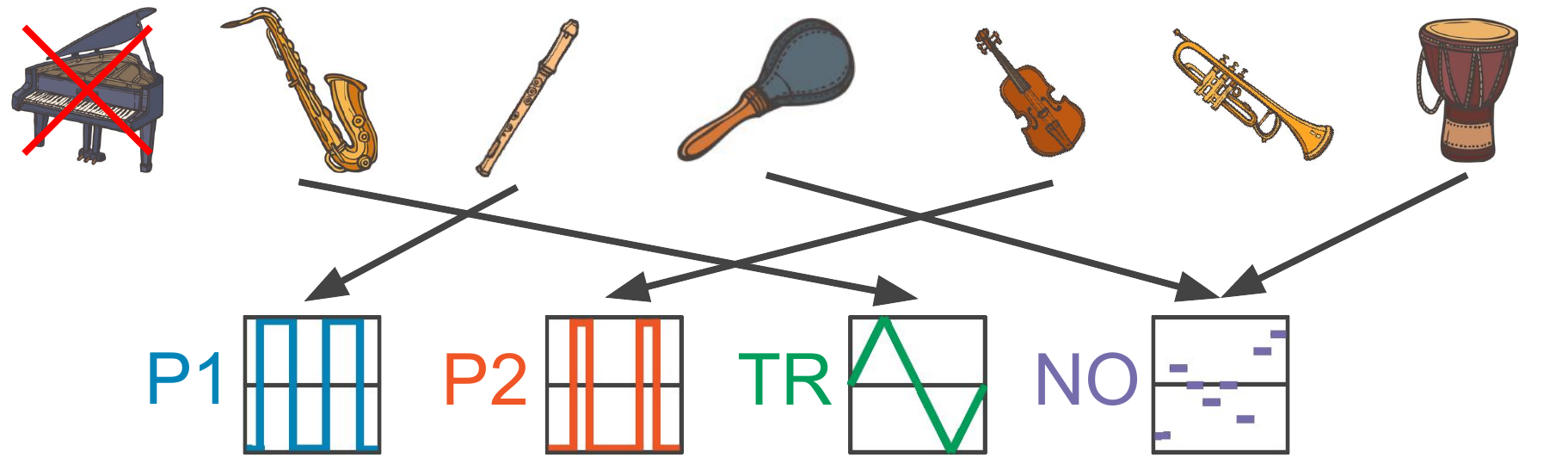}
    \caption{Illustration of our mapping heuristic used to enable transfer learning from Lakh MIDI to NES-MDB. We identify monophonic instruments from the arbitrary ensembles in Lakh MIDI and randomly assign them to the fixed four-instrument ensemble of NES-MDB.}
    \label{fig:lakh2nes}
\end{figure}

Here we describe our protocol for mapping Lakh MIDI data into a score suitable for the four monophonic instruments of the NES ensemble. For a given example from Lakh MIDI, we first identify all of its monophonic melodic instruments (skipping the example if it has no such instruments). 
Then, we filter out instruments which fall outside of the range of MIDI notes that the NES ensemble is capable of producing (\Cref{sec:nesens}). 
We randomly assign these instruments to the three melodic instruments of the NES (P1/P2/TR) (\Cref{fig:lakh2nes}). 
Because there are a variable number of instruments in each Lakh MIDI example, there are potentially many possible assignments. 
Hence, we output multiple examples for each input Lakh MIDI example. 

In addition to this strategy for melodic instruments, we also design a strategy for mapping percussive instruments in Lakh into the percussive noise instrument of the NES ensemble. 
We first identify percussive instruments in each Lakh MIDI example. 
Then, 
each individual percussive voice (e.g.,~snare drum, hi-hat) is randomly assigned to a noise ``type'' ($1$--$16$), 
emulating how the noise instrument is used by human composers to encode syncopated rhythms.

From the $175$k MIDI files in Lakh MIDI, 
our mapping procedure produces $775$k examples suitable for performance by the NES ensemble. 
It is straightforward to imagine similar mapping procedures for other ensembles (e.g.,~string quartet, vocal choir), 
and thus it is possible that music generation research in other domains could reuse this procedure to enable transfer learning.
%We will release all code for our mapping pipeline upon publication.

\section{Experiments}
\label{sec:experiments}

We first conduct an experiment to train Transformer-XL~\cite{dai2019transformer} on our event representation (\Cref{sec:eventbased}) of NES-MDB. 
We train the model on excerpts from the training data of $512$ events; each excerpt represents around $9$ seconds of music on average. 
Because of the recurrent attention mechanism in Transformer-XL, the model effectively has access to twice this length in its history.

We use the smaller configuration of Transformer-XL which has $12$ attention layers each with $8$ heads. 
The learning rate $2\mathrm{e}{-4}$ used to train this model on text was found to be too high for our musical application, 
so we lowered it to $2\mathrm{e}{-5}$. 
Training was stopped when the performance of the model on the validation data stopped improving. 
We trained the model using four NVIDIA Titan X GPUs with minibatches of size $30$, 
and it reached its early stopping criteria in less than a day.\footnote{
Full hyperparameter description and pre-trained models: \\ \url{https://github.com/chrisdonahue/LakhNES}}

\subsection{Data augmentation and pre-training}

To improve the performance of our model further, we employed standard music data augmentation methods as well as ones which we developed specifically for the multi-instrumental setting:

\begin{enumerate}
    \item (Standard) Transpose melodic voices by a random number of semitones between $-6$ and $5$ (inclusive).
    \item (Standard) Adjust the speed of the piece by a random percentage between $\pm 5$\%.
    \item Half of the time, remove a random number of instruments from the ensemble (leaving at least one).
    \item Half of the time, shuffle the score-to-instrument alignment for the melodic instruments only (e.g.,~TR performs P2's part).
\end{enumerate}

Finally, we experimented with pre-training our model on the Lakh MIDI dataset mapped to the NES ensemble (\Cref{sec:lakh2nes}). 
To conduct this experiment, we first split the Lakh data into training and validation subsets. 
We then trained the model for a week on the training set (with data augmentation) and monitored performance on the validation set. 
Because of the extreme size of the dataset, the model only completed four epochs of training. 
Even after a week, the model was underfitting the training data (validation performance was still improving).
We then fine-tuned this pre-trained model on the NES-MDB training data, again performing early stopping based on the validation performance. 
Both our pre-training and fine-tuning experiments use the same hyperparameters outlined in the previous section.

\subsection{Baselines}

We also measure the performance of competitive baselines on our event-based representation of NES-MDB. 
Our simplest baselines consist of $n$-gram models, i.e.,~statistics gathered directly from the training data of how often certain length-$n$ sequences appear. 
Specifically, we build unigram ($1$-gram) and $5$-gram models, using backoff for the latter to provide a likelihood for $5$-grams which are not present in the training data. 
We also compare to an LSTM~\cite{hochreiter1997long} recurrent neural network, which is a popular model for music generation. 
Our LSTM is configured so that it has approximately the same number of parameters as our Transformer-XL model ($1$ layer, $3072$ units).

\section{Quantitative analysis}

\begin{table}[t]
\centering
\begin{tabular}{lrrr}
\toprule
Model & Params & Epochs & Test PPL \\
\midrule
Random & $0$ & $0$ & $631.00$ \\
Unigram & $631$ & $1$ & $198.14$ \\
$5$-gram & $9$M & $1$ & $37.25$ \\
LSTM~\cite{hochreiter1997long} & $40$M & $18$ & $14.11$ \\
~~+Data augmentation &  & $35$ & $12.64$ \\
Transformer-XL~\cite{dai2019transformer} & $41$M & $76$ & $3.50$ \\
~~+Data augmentation &  & $350$ & $2.74$ \\
~~+Pre-train (LakhNES) &  & $250$ & $\mathbf{2.46}$ \\
\bottomrule
\end{tabular}
\caption{Quantitative performance of various models trained on the event-based representation ($631$ event types) of NES-MDB. \emph{Params} indicates the number of parameters of each model. \emph{Epochs} is the number of data epochs the model observed before early stopping based on the validation data. \emph{Test PPL} represents the perplexity of the model on the test data, i.e.,~the exponentiation of its average negative log-likelihood on the test data. A lower perplexity indicates that the model better fits this unseen data.}
\label{tab:quant}
\end{table}

We report the \emph{perplexity} (PPL) of each model on the test set in \Cref{tab:quant}. 
Perplexity is calculated by first averaging the negative log-likelihood of each model across the test data, 
then exponentiating the average, i.e.,~$e^{\frac{1}{N}\sum_{i=1}^{N}-\log q_i}$, where $q_i$ is the likelihood assigned by a given model to the $i$-th event.
A lower perplexity on the test set indicates that a model is a good fit for unseen data, and hence increases our confidence in its ability to generate new music.

We find that Transformer-XL dramatically outperforms both the $n$-gram and LSTM baselines on the NES-MDB event-based task (PPL of $3.5$ vs.~$37.2$ and $14.1$ respectively). 
Data augmentation improves the performance of both the LSTM and Transformer-XL (by $10$\% and $22$\% respectively), and also increases the number of epochs before the models overfit.
%the models observe before overfitting. 
We observe that LakhNES (Transformer-XL pre-trained on Lakh MIDI and fine-tuned on NES-MDB with augmentation), 
achieves $10$\% better performance than training with data augmentation alone.
% \textbf{We observed that LakhNES combining data augmentation with our Lakh MIDI training procedure improved the performance of Transformer-XL by an \emph{additional} $10$\%}. 

\begin{figure}[t]
    \centering
    \includegraphics[width=0.99\columnwidth]{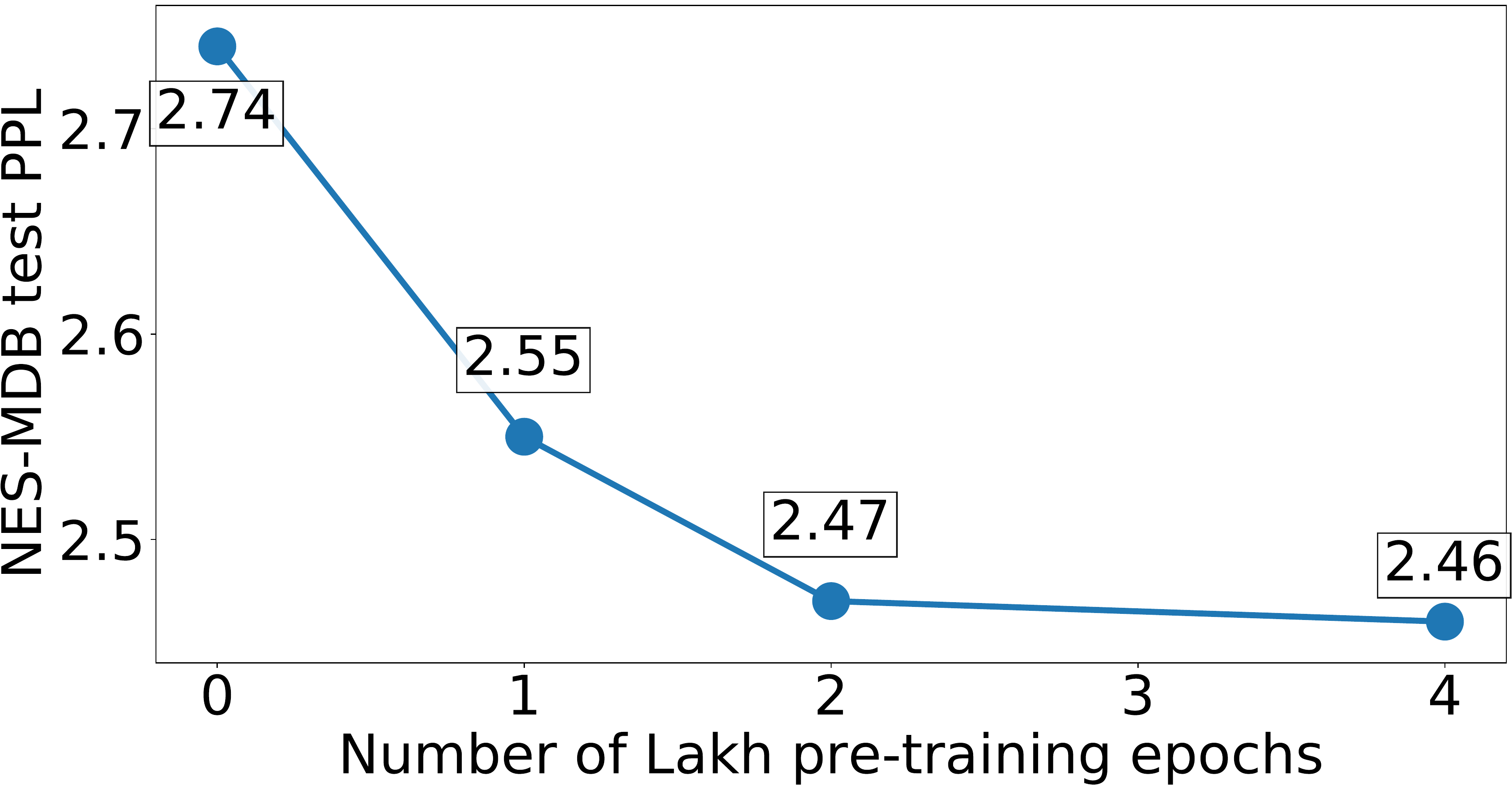}
    \caption{Measuring the performance improvement when doubling the amount of Lakh MIDI pre-training before fine-tuning Transformer-XL on NES-MDB. Each datapoint represents the result of a fine-tuning run starting from $0$, $1$, $2$, or $4$ epochs of Lakh MIDI pre-training. Additional amounts of pre-training appear to improve performance, though with diminishing returns.}
    \label{fig:pretrend}
\end{figure}

We also conduct an experiment to measure the performance effect of using different amounts of Lakh MIDI pre-training before fine-tuning on NES-MDB.
Specifically, 
we measure the performance on the NES-MDB fine-tuning task after $1$, $2$, and $4$ epochs of Lakh MIDI pre-training. 
We plot the test PPL of each model after fine-tuning in \Cref{fig:pretrend}. 
%We see that 
The results agree with our expectation that increasing the amount of pre-training improves the fine-tuned model's performance, 
though with diminishing returns. 

\section{User study}

While perplexity is a useful quantitative metric for model comparison, 
it is not necessarily correlated with human judgements. 
Since we ultimately seek models which produce music that is convincing to humans, we conduct two user studies on Amazon Mechanical Turk to compare the performance of various models. 
In both of our user studies we compare four models (rows 3, 5, 7, 8 from~\Cref{tab:quant}): (1) $5$-gram model, (2) LSTM trained with data augmentation, (3) Transformer-XL trained with data augmentation (TXL), and (4) Transformer-XL with data augmentation and Lakh MIDI pre-training (LakhNES).
%Random examples from all of our methods can be heard at the bottom of our sound examples page: \url{https://bit.ly/2Z1FDGh}~~~. 

\subsection{Turing test}
\label{sec:turi}

This study seeks to determine the ability of humans to distinguish between real (human-composed) and fake (computer-generated) chiptunes in a ``Turing test'' setting. 
We present human judges with pairs of examples where one example is real and the other fake, and ask them to identify the real example between the two.

We first amass collections of $5$-second audio clips from all of our methods and from the real data by selecting random slices from the variable-length music. 
Then, we create pairs of examples where one example is real and the other fake (randomly chosen from our four methods). 
Given that Mechanical Turk studies are notoriously noisy, we also create control pairs where the fake data comes from a random model (i.e.,~we generate ``music'' by selecting events uniformly at random---row 1 in \Cref{tab:quant}).

\begin{figure}[t]
    \centering
    \includegraphics[width=0.99\columnwidth]{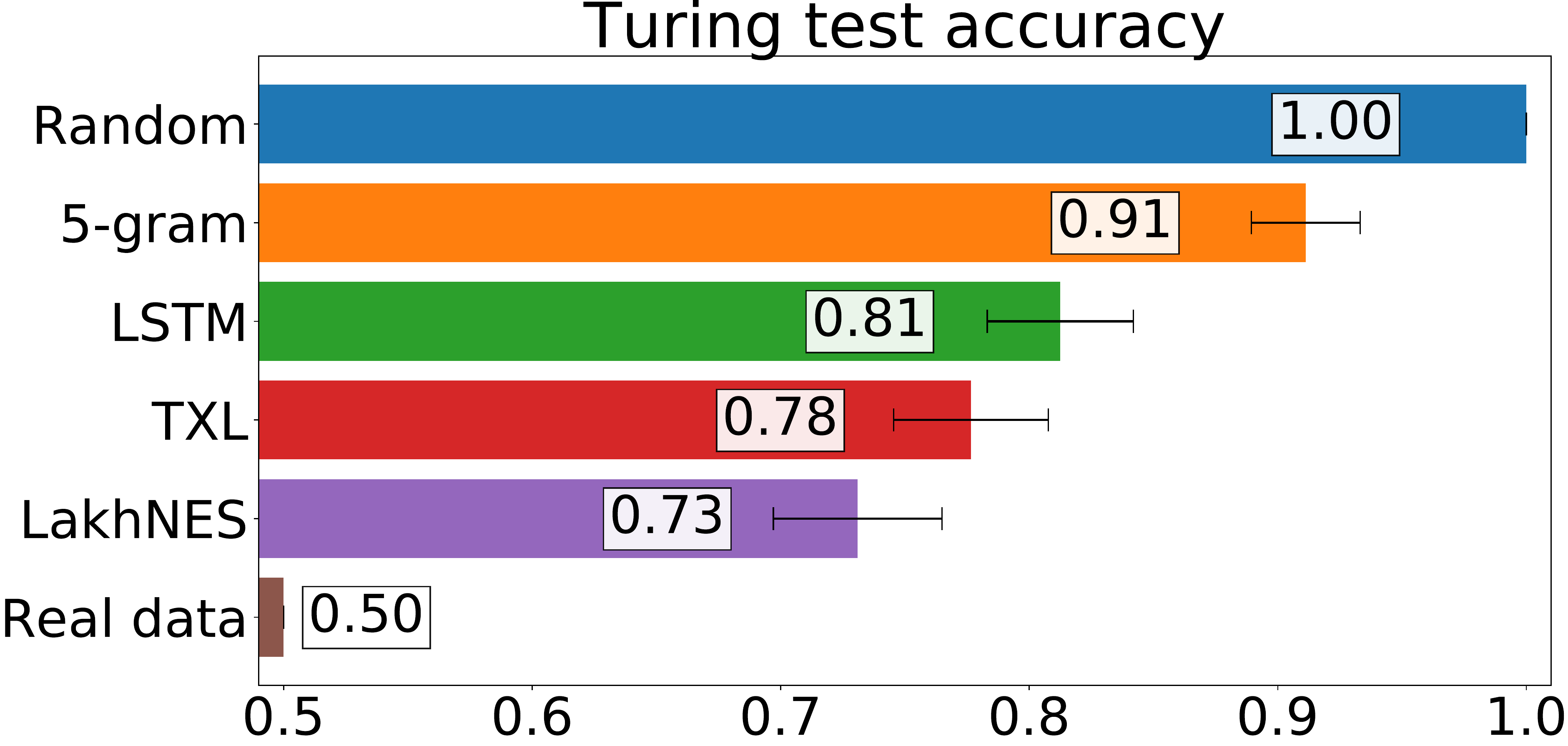}
    \caption{Human accuracy at distinguishing computer-generated examples from human-composed ones (error bars are standard error). Users were presented with pairs of clips (one human, one computer) and tasked with identifying which was composed by a human. Random examples are used as a control and we filtered annotators with accuracy less than $1$ on those pairs. A lower accuracy is better as it indicates that the annotators confused a particular model with the real data more often.}
    \label{fig:turi}
\end{figure}

We ask human judges to annotate $800$ batches each consisting of $10$ randomly-ordered pairs, 
where fake data in $2$ of the pairs came from the control set and fake data in $8$ of the pairs came from our four methods.
For their judgments to be included in our results, 
workers were required to complete at least $3$ batches and achieve $100\%$ accuracy on the $6$ control examples in those batches---a worker answering randomly would only be included $1.6$\% of the time.
%We found this control-based filtering to greatly improve the signal in our studies, and recommend similar strategies for others employing Mechanical Turk for music-oriented tasks.
After filtering, each method was evaluated around $180$ times.
We report accuracy in \Cref{fig:turi}.

In this setting, 
a lower accuracy indicates that a given model's results sound more human-like, 
because they were incorrectly identified as human-composed more often.
An ideal generative model would achieve $50$\% accuracy (although it is possible in theory to generate music which sounds ``more human'' than human-composed music).
We find that LakhNES (Transformer-XL with pre-training) was mistakenly identified as human more often than both our $5$-gram model ($p<.0001$ by \emph{t}-test with normal approximation) and our LSTM ($p=.07$). 
It also outperformed Transformer-XL without pre-training, but the difference was not statistically significant 
%but with low confidence 
($p=.32$).

Overall, these results suggest that there is still a sizable gap between human-composed and computer-generated chiptunes.
%We suspect that results would be even less favorable for the generative methods if clips longer than $5$ seconds were used.
Subjectively speaking, 
we feel that the melodies and harmonies produced by LakhNES are promisingly human, 
but its inability to maintain rhythmic consistency is often a dead giveaway in a Turing test.
We suspect that our model could be improved by using a beat-based event representation, however the current model can be bootstrapped with human-specified rhythmic material to manually address rhythmic consistency issues (\Cref{sec:pilot}).

\subsection{Preference test}
\label{sec:pref}

In addition to our Turing test, 
we also conduct a preference-based user study, 
given that human-ness is not necessarily a predictor of general preference. 
We present human judges with pairs of examples from two different methods, and ask them which of the two they ``prefer''.

\begin{figure}[t]
    \centering
    \includegraphics[width=0.99\columnwidth]{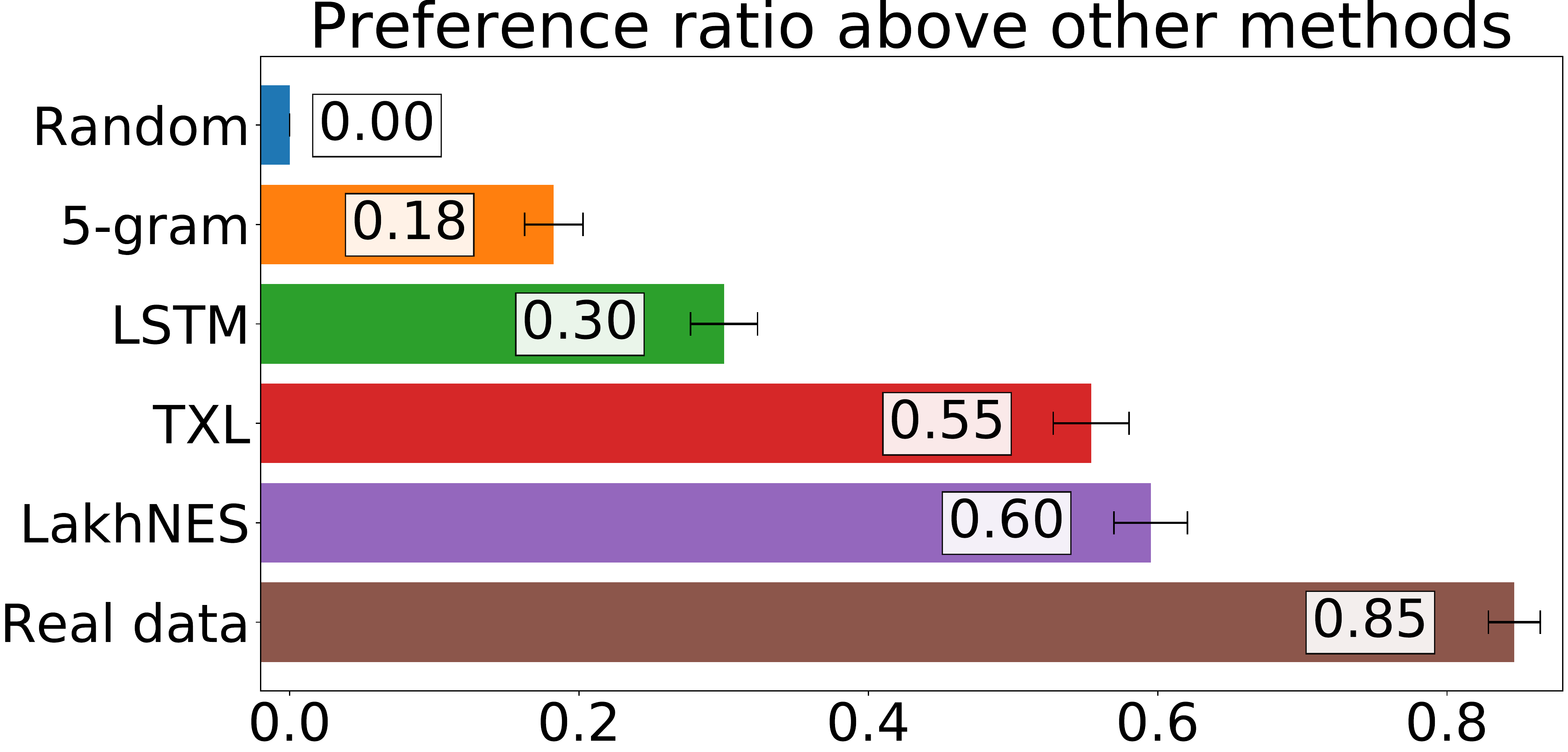}
    \caption{Proportion of comparisons where humans preferred an example from each model over an example from another random model (error bars are standard error). Users were presented with pairs of clips from different methods and asked which they preferred. Pairs of random data and human-composed clips are used as a control and we filtered annotators who preferred random. A higher ratio is better as it indicates that the annotators preferred results from that method more often than another.}
    \label{fig:pref}
\end{figure}

Here we construct pairs of $10$-second clips from two different (randomly-chosen) methods. 
These clips are twice as long as those used in~\Cref{sec:turi} allowing longer-term structure to influence preference decisions.
As in our Turing test, we construct randomly-ordered batches consisting of $10$ pairs. 
In each batch, $8$ of the pairs are created by sampling two methods without replacement from a set of five (four computer-generated and the real data), 
while $2$ pairs always compare randomly-generated clips to real data (control). 
We ask human judges to assign preference to these batches, 
filtering out workers who even once indicated that they preferred random examples to real data.
After filtering, each of the five methods was involved in around $400$ comparisons in total. 
We report the ratio of ``wins'' for each method in \Cref{fig:pref}, i.e.,~the proportion of times a method was preferred over any of the other four.

We find that LakhNES outperforms all other generative methods, 
though is preferred significantly less often than the real data. 
% 6% of the time for the 5-gram
Human judges preferred chiptunes generated by LakhNES over the real data in $26$\% of comparisons 
(vs. only $10$\% of the time for the LSTM). 
We find this to be a promising indicator of Transformer's potential on this task.

\section{Pairing LakhNES with humans}
\label{sec:pilot}

In addition to generating chiptunes from scratch,
LakhNES can be used for a number of tasks to assist human composers.
%due to the convenience of the event-based representation. 
% Due to the convenient event-based factorization of our music representation, 
% there are a number of use cases for which LakhNES can immediately be used to assist human composers. 
For example, LakhNES can be ``primed'' on human-composed material and then asked to continue the material, 
providing a method for composers to quickly expand on their ideas. 
Composers can also provide fixed rhythmic material and use LakhNES to generate the rest of the score. 
We explore these use cases in our sound examples: \url{https://chrisdonahue.com/LakhNES}~.
When generating all of our sound examples (besides those in our user study),
we found that limiting the entropy of the model by using a sampling temperature of $.95$ and top-$k$ sampling~\cite{fan2018hierarchical} with $k=32$ improved results qualitatively.

\section{Conclusion}

In this paper we presented LakhNES, a method for learning to generate multi-instrumental music. 
We developed an event-based representation suitable for this task.
%, a first in music generation research.
%which has not previously been explored by the music generation community. 
Training powerful language models on this representation results in compelling multi-instrumental music generation. 
We show that we can further improve results both quantitatively and qualitatively by pre-training on a cross-domain dataset. 
%This cross-domain pre-training procedure could be useful for overcoming data limitations for other musical domains. 
LakhNES can be used to both generate chiptunes from scratch and collaborate with human composers.
%in a number of different ways. 
%The chiptunes produced by our system could potentially provide a procedural soundtrack for new games developed by the burgeoning retro gaming community.

\section{Acknowledgements}

Thanks to Cheng-Zhi Anna Huang, Cheng-i Wang, and Jennifer Hsu for helpful discussions regarding this work.
This work was supported by UC San Diego's Chancellors Research Excellence Scholarship program.
GPUs used in this research were donated by NVIDIA.

% For bibtex users:
\bibliography{ISMIR2019template}

\end{document}